\documentclass{iaus}
\usepackage{graphicx}

\title[On TP-AGB stars and the mass of galaxies] 
{On TP-AGB stars and the mass of galaxies}

\author[Bruzual]{Gustavo Bruzual A.$^1$}

\affiliation{$^1$CIDA, AP 264, M\'erida, Venezuela\break email: bruzual@cida.ve}

\pubyear{2006}
\volume{241}
\pagerange{119--126}
\date{?? and in revised form ??}
\jname{Stellar Populations as Building Blocks of Galaxies}
\editors{A. Vazdekis \& R. Peletier, eds.}
\begin{document}

\maketitle

\begin{abstract}
Recent calculations of evolutionary tracks of TP-AGB stars of different mass
and metallicity by \cite{PM07} have been incorporated in the Bruzual \& Charlot
evolutionary population synthesis models. The mass of the stellar population in HUDF galaxies at
$z$ from 1 to 3 determined from fits to the spectro-photometric data of these galaxies
is 50 to 80\% lower than the mass determined from the BC03 models.
The ages inferred for these populations are, with exceptions, 40 to 60\% of the BC03 estimates.
\end{abstract}

\firstsection 

\section{Introduction}

\begin{figure}[!ht]
\begin{center}
\includegraphics[scale=0.70]{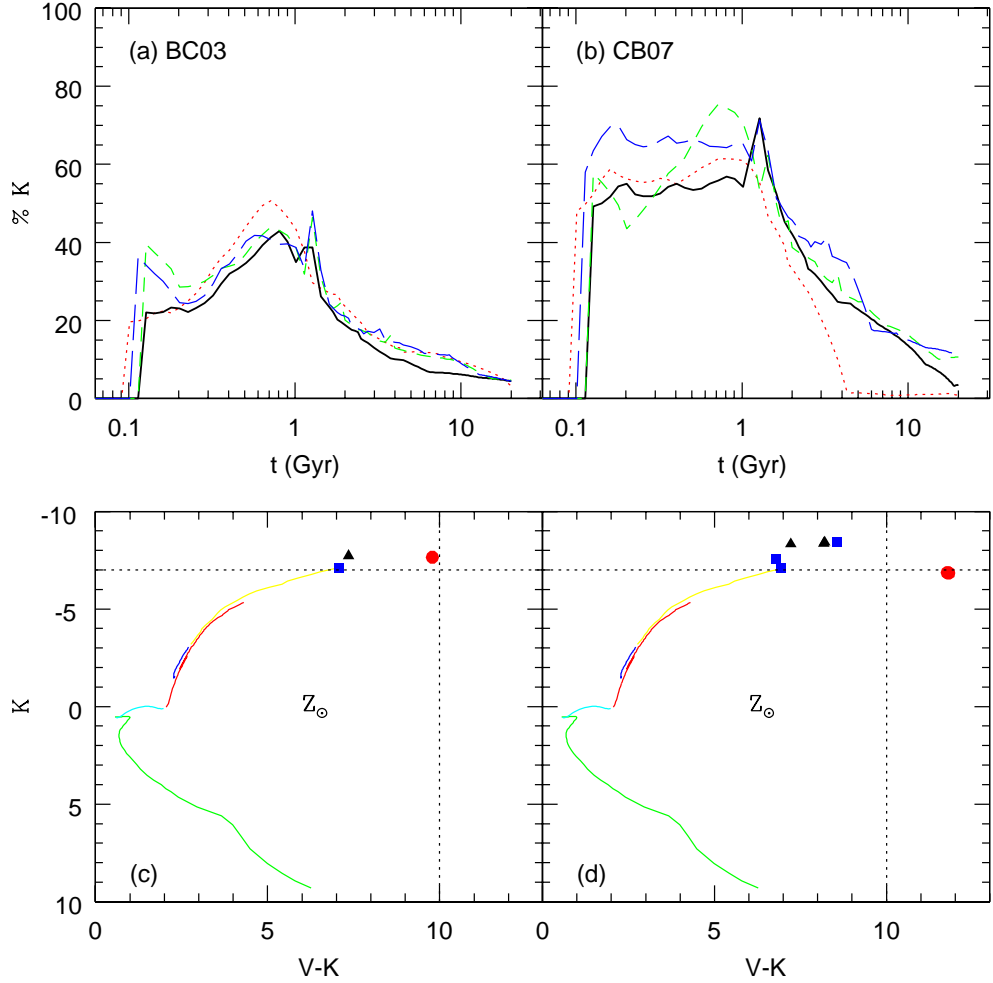}
\end{center}
\caption{
  Top frames: time dependence of the fraction of the galaxy luminosity
  emitted in the K-band according to {\it (a)} the BC03 and {\it (b)}
  the CB07 SSP models. Different lines correspond to different metallicities
  of the stellar population, as follows: 
  $Z_\odot$ (black solid line),
  $0.4 \times Z_\odot$ (green long-dashed line),
  $0.2 \times Z_\odot$ (blue short-dashed line),
  $2.5 \times Z_\odot$ (red dotted line).
  All models shown in this figure were computed for the Padova 1994 tracks
  assuming the \cite{CHAB03} IMF.
  Figures drawn in color are available in the electronic version of this paper.
  Bottom frames: 1.3 Gyr $Z_\odot$ isochrone computed using {\it (c)} the BC03 and
  {\it (d)} the CB07 models. The lines representing the MS, SGB, RGB, HB,
  and AGB are identical in both cases, and correspond to the Padova 1994 tracks.
  The symbols represent the O-rich (blue squares), the C-rich (black triangles),
  and the superwind (red circles) phases of the TP-AGB.
  The dotted lines are drawn to guide the eye.
}
\end{figure}

In the oral presentation of this paper I gave a summary of the current
status of various high resolution stellar spectral libraries and their
use in population synthesis models. A written version of this summary
can be found in a series of conference papers appeared in recent
years, (Bruzual 2004, 2005, 2006) and will not be repeated here for
reasons of space.
In this paper I will concentrate in the second part of my talk which
versed on the r\^ole of the thermally pulsing asymptotic giant branch
(TP-AGB) phase of stellar evolution on the integrated properties of stellar
populations.

It has been pointed out by several authors, e.g. \cite{CM06}, \cite{KG07},
that the estimates of the age and mass of the stellar population present in a galaxy
depend critically on the ingredients of the stellar population model
used to fit the galaxy spectrum. \cite{CM06} have shown that the treatment 
of the TP-AGB phase of stellar
evolution is a source of major discrepancy in the determination
of the spectroscopic age and mass of high-z $(1.4<z<2.7)$ galaxies.
The mid-UV spectra of these galaxies indicate ages in the range from
0.2-2 Gyr, at which the contribution of TP-AGB stars in the rest-frame
near-IR sampled by Spitzer is expected to be at maximum. \cite{CM06} find that
in general the \cite{CM05} models (M05 hereafter) provide better fits to
the observations than the \cite{BC03} models, hereafter BC03,
and other models available in the literature, and indicate systematically lower ages
and, on average, 60\% lower masses for the stellar populations sampled in these galaxies.
According to \cite{CM06} the source of this discrepancy is primarily a consequence
of the different treatment of the TP-AGB phase in the evolutionary models.

\begin{figure}[!ht]
\begin{center}
\includegraphics[scale=0.70]{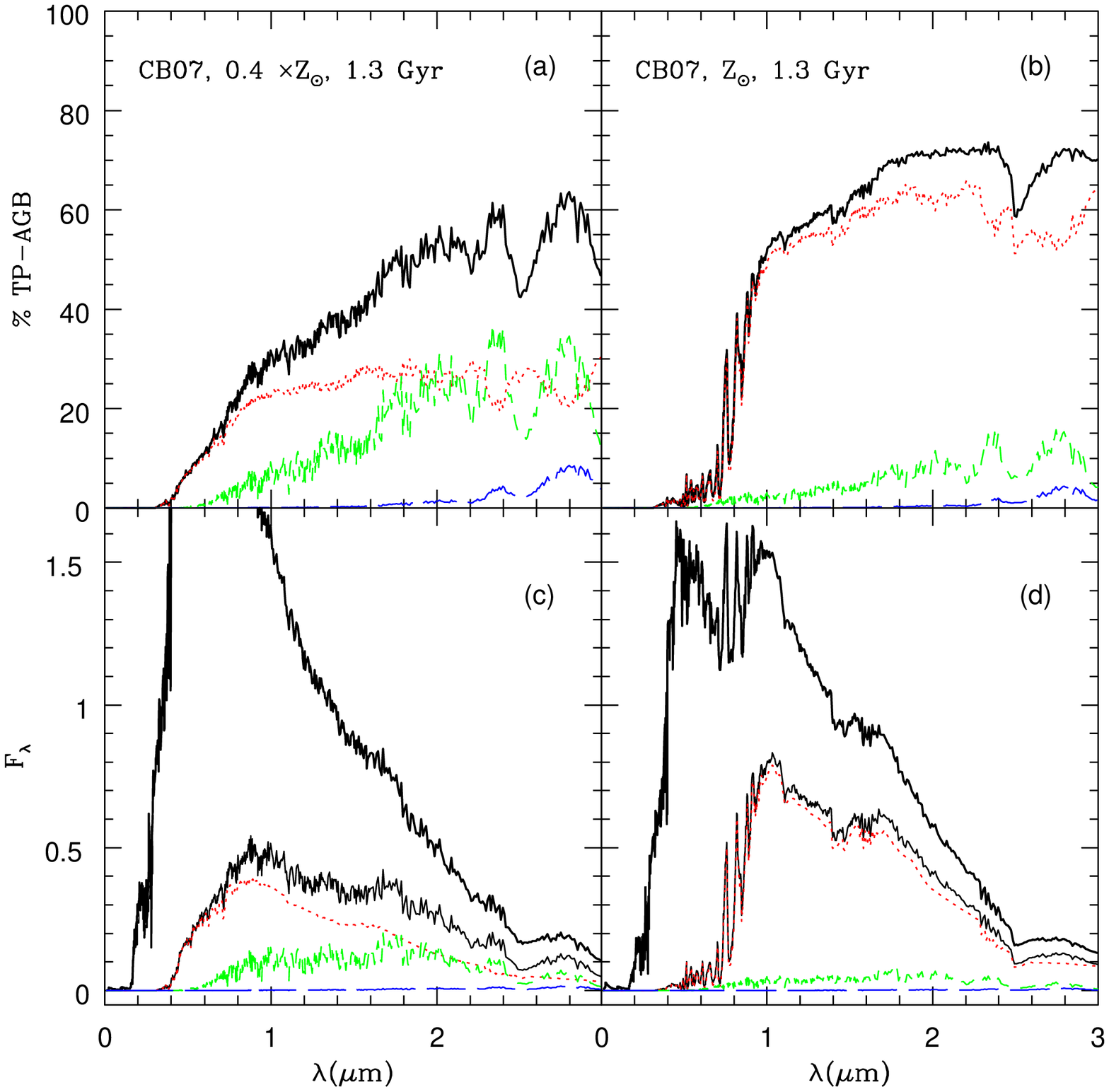}
\end{center}
\caption{
  Top frames: Percentage of the luminosity of the 1.3 Gyr SSP emitted by TP-AGB
  stars as a function of wavelength (heavy solid-line) according to the CB07
  models for {\it (a)} $Z = 0.4 \times Z_\odot$, and {\it (b)} $Z = Z_\odot$.
  The contribution of the O-rich TP-AGB stars (red dotted-line),
  the C-rich TP-AGB stars (green short-dashed line),
  and the TP-AGB stars in the superwind phase (blue long-dashed line) is
  also indicated. Figures drawn in color are available in the
  electronic version of this paper.
  Bottom frames: spectral energy distribution of the 1.3 Gyr SSP as a function
  of wavelength (heavy solid-line) according to the CB07 models for
  {\it (c)} $Z = 0.4 \times Z_\odot$, and {\it (d)} $Z = Z_\odot$.
  The contribution of all the TP-AGB stars (thin solid-line),
  the O-rich TP-AGB stars (red dotted-line),
  the C-rich TP-AGB stars (green short-dashed line),
  and the TP-AGB stars in the superwind phase (blue long-dashed line) is
  also indicated.
}
\end{figure}

\cite{PM07} have recently concluded new calculations of the TP-AGB evolutionary
phase for stars of different mass and metallicity. The evolution of the stars is
now computed accounting for the changes in the chemical composition of the envelopes.
As a consequence of this prescription, the signature of TP-AGB stars around 1 Gyr,
i.e the red color of the integrated stellar population, becomes more relevant than
in previous computations. The full implementation of the new stellar libraries and the 
\cite{PM07} TP-AGB evolutionary tracks in population synthesis models is in preparation
by Charlot \& Bruzual (2007, CB07 hereafter).
All the BC03 and CB07 SSP models shown in this paper have been computed
for the \cite{CHAB03} IMF and the \cite{WES02} stellar library.
Fig. 1 shows the time dependence of the fraction
of the galaxy luminosity emitted in the K-band according to the BC03
and CB07 models.
From the top frames of Fig. 1 it is apparent that the TP-AGB stars in the CB07
models contribute at least a factor of two more light in the $K$-band than in the
BC03 models, which use a different prescription for the TP-AGB evolution (see BC03
for details).  
At maximum, the TP-AGB contributes close to 70\% of the K-light in the CB07 model
but only 40\% in the BC03 model. The peak emission in the BC03 model occurs
at around 1 Gyr whereas in the CB07 model it stays high and close to constant
from 0.1 to 1 Gyr.
To explore the cause of this difference, I plot in the bottom frames of Fig. 1
the 1.3 Gyr $Z = Z_\odot$ isochrone for both sets of models.
This age corresponds to the spike seen in the $Z = Z_\odot$ line in the top frames of
this figure.
Whereas the \cite{PM07} tracks used in CB07 account for 9 evolutionary stages in the
TP-AGB (three in the O-rich phase, three in the C-rich phase, and three in the superwind
phase), the BC03 models include only 1 evolutionary stage on each of these phases.
The TP-AGB stars in the CB07 isochrone are close to 1 magnitude brighter in $K$
and reach values of $V-K$ several magnitudes redder than their counterparts in BC03.
The evolutionary rate is such that the total number of TP-AGB stars present in the
CB07 1.3 Gyr isochrone is 4 times larger than the number of these stars present
in the BC03 models. The TP-AGB stars represent 0.016\% of the total number of
stars present in this population at this age in the CB07 model, but only 0.004\% in
the BC03 model.
The net effect of all these factors is the increased contribution shown in Fig. 1.

\begin{figure}[!ht]
\begin{center}
\includegraphics[scale=0.70]{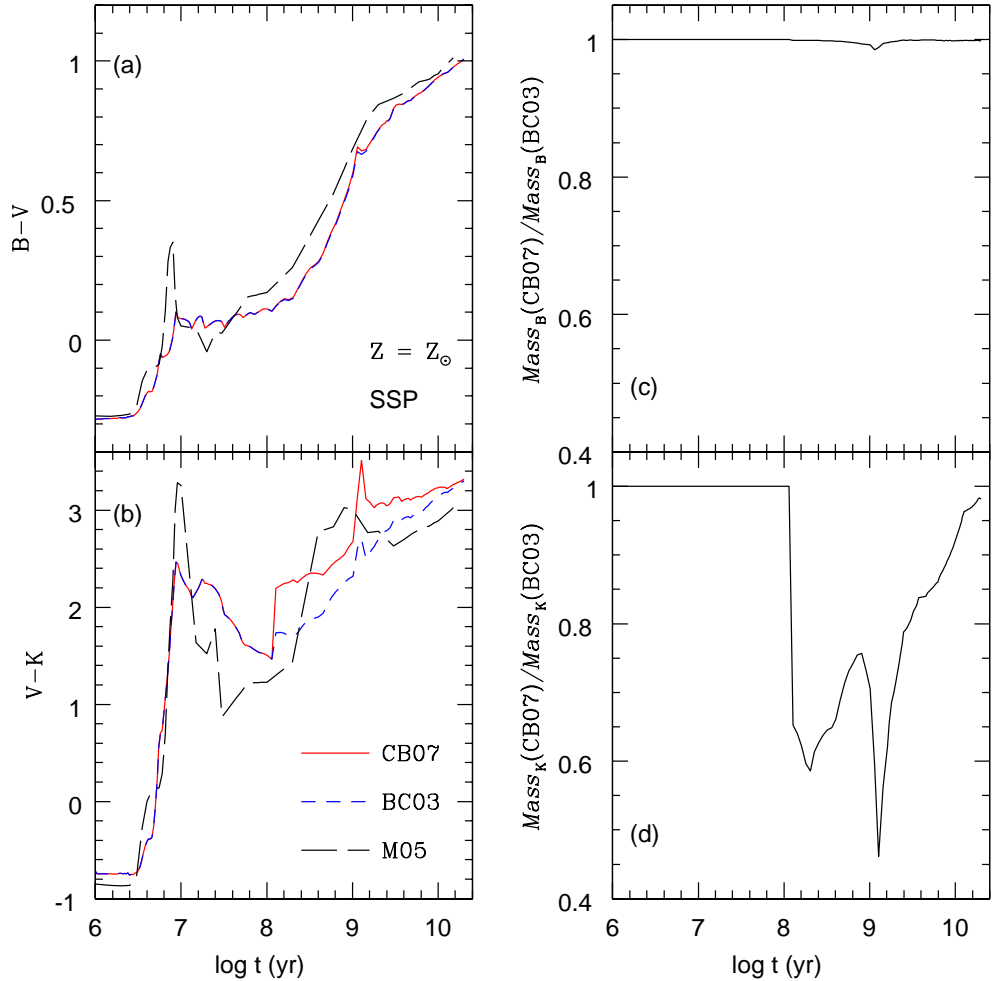}
\end{center}
\caption{
Comparison of {\it (a)} the B-V and {\it (b)}  V-K color evolution of
the CB07 models (red solid line)
with the BC03 (blue short-dashed line) and the M05 (black long-dashed line) models.
Figures drawn in color are available in the electronic version of this paper.
{\it (c)} and {\it (d)}: ratio of the stellar mass determined from the CB07 and the
BC03 models for a given $B$ and $K$-band galaxy luminosity, respectively.
All models shown in this figure correspond to a $Z = Z_\odot$ SSP computed for the
\cite{CHAB03} IMF.
}
\end{figure}

\begin{figure}[!ht]
\begin{center}
\includegraphics[scale=0.7]{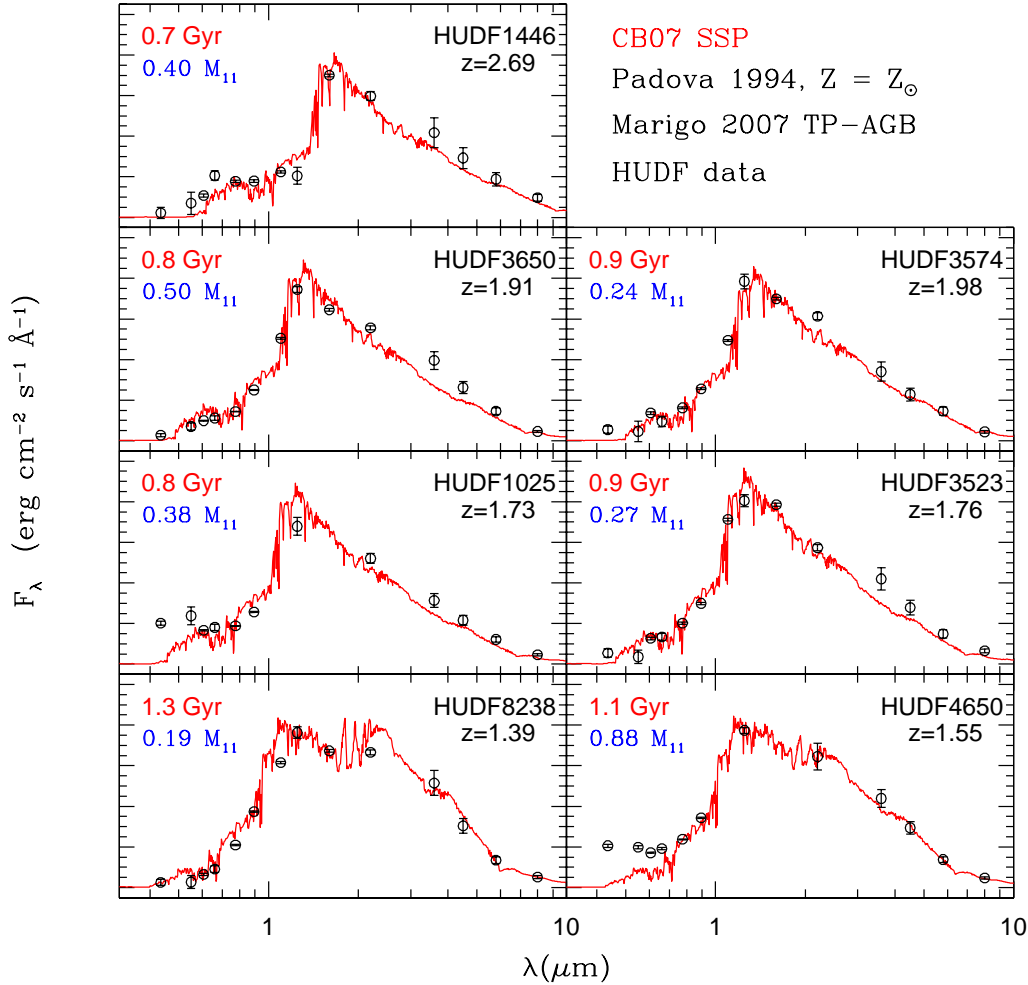}
\end{center}
\caption{
Observed spectral energy distribution (symbols with error bars) of 7 HUDF
galaxies taken from Table 1 of \cite{CM06} shown together with the best
fitting CB07 $Z=Z_\odot$, Chabrier IMF SSP model (red solid-line).
The galaxy HUDF identification
number, redshift, age, and mass (in units of $10^{11}$ M$_\odot$) are
given inside each frame and listed in Table 1.
For clearness, after determining the galaxy mass the spectra were forced to
share the same vertical scale.
Figures drawn in color are available in the electronic version of this paper.
}
\end{figure}

\begin{table}\def~{\hphantom{0}}
  \begin{center}
  \caption{Spectroscopic age and mass$^3$ of HUDF galaxies}
  \label{tab:kd}
  \begin{tabular}{lcccccccccc}\hline
             &      & ~~~ &  CB07$^1$ & CB07$^1$ & ~~~ & BC03$^2$ & BC03$^2$ ~~~ & M06$^2$ & M06$^2$ \\
        HUDF &  z   & ~~~ & t(Gyr)& M$_{11}$& ~~~ & t(Gyr)& M$_{11}$ ~~~ & t(Gyr)& M$_{11}$ \\\hline
	1446 & 2.69 & ~~~ &  0.72 & 0.40 & ~~~ & 2.30 & 0.80 ~~~ & 0.70 & 0.40 \\
	3650 & 1.91 & ~~~ &  0.81 & 0.50 & ~~~ & 2.30 & 1.60 ~~~ & 0.50 & 0.65 \\
	1025 & 1.73 & ~~~ &  0.81 & 0.38 & ~~~ & 1.70 & 1.25 ~~~ & 1.70 & 1.00 \\
	8238 & 1.39 & ~~~ &  1.28 & 0.19 & ~~~ & 2.30 & 1.13 ~~~ & 1.00 & 0.46 \\
	3574 & 1.98 & ~~~ &  0.90 & 0.24 & ~~~ & 2.60 & 0.53 ~~~ & 2.30 & 0.56 \\
	3523 & 1.76 & ~~~ &  0.90 & 0.27 & ~~~ & 2.30 & 1.00 ~~~ & 1.40 & 0.60 \\
	4650 & 1.55 & ~~~ &  1.10 & 0.88 & ~~~ & 1.70 & 3.50 ~~~ & 1.70 & 2.20 \\\hline
  \end{tabular}
  \begin{tabular}{l}
   ~~~$^1$ Determined from fitting a $Z = Z_\odot$, Chabrier IMF SSP to the magnitudes~~
  \end{tabular}
  \begin{tabular}{l}
   ~~~~~~~~~~in Table 1 of \cite{CM06}.~~~~~~~~~~~~~~~~~~~~~~~~~~~~~~~~~~~~~~~~~~~~~~~~~~~~~~~~~~~~
  \end{tabular}
  \begin{tabular}{l}
   ~~$^2$ Values read from Fig. 4 of \cite{CM06} assuming $E(B-V) = 0$.
  \end{tabular}
  \begin{tabular}{l}
   ~~$^3$ The galaxy mass M$_{11}$ is given in units of $10^{11}$ M$_\odot$.~~~~~~~~~~~~~~~~~~~~~
  \end{tabular}
 \end{center}
\end{table}

Fig. 2 shows the contribution of TP-AGB stars to the spectral energy
distribution of the 1.3 Gyr SSP as a function of wavelength according to the CB07
models for $Z = 0.4 \times Z_\odot$, and $Z = Z_\odot$.
The relative contribution of the O-rich vs. the C-rich TP-AGB stars depends
critically on the assumed stellar metallicity.
Fig. 3 compares the $B-V$ and $V-K$ color evolution of the CB07, BC03 and M05 models.
In $B-V$ the CB07 and BC03 models are identical at all ages. At early and late ages
both sets of models have the same $V-K$ color, but at intermediate ages the CB07 models
are considerably redder than the BC03 models. At late ages, the BC03 and CB07 models
match very well the observations of nearby early-type galaxies, whereas the M05 models
are too blue. The stellar mass determined from the $K$-band luminosity using the
CB07 model can be up to 50\% lower than the mass determined from the BC03
model at the same age. The BC03 and CB07 models will predict the same galaxy mass if
the $B$-band luminosity is used.

Fig. 4 compares the observed spectral energy distribution of 7 HUDF galaxies with the
best fitting CB07 $Z=Z_\odot$, Chabrier IMF SSP model. The fits look quite reasonable and can be
improved assuming different star formation histories or dust content.
The galaxy HUDF identification
number, redshift, and derived age and mass (in units of $10^{11}$ M$_\odot$) are
given inside each frame and listed in Table 1.
As expected from the previous discussion, the mass and age of the stellar
population determined from the level of the IR flux observed in these galaxies are
considerably lower than the values determined by \cite{CM06} from the BC03 (and M05) models.
The masses derived from the CB07 models are 50 to 80\% lower than the BC03 masses.
The ages derived from the CB07 models are, with exceptions, 40 to 60\% of those derived from BC03.

\acknowledgements

I thank Paola Marigo and Leo Girardi for providing their calculations of the TP-AGB 
evolutionary phase ahead of publication, and St\'ephane Charlot for allowing me
to show results of a joint paper in preparation.

\begin{discussion}

\discuss{Kroupa}{How could your results on the galaxies (their ages and masses) change
if the IGIMF is steeper than the standard IMF?.}

\discuss{Bruzual}{If the number of intermediate mass stars is reduced as a consequence of having
a different IMF, we expect a lower contribution of TP-AGB stars to the galaxy luminosity. I would
guess that the estimates of the galaxy mass will be somewhat higher than for the standard IMF. The
age estimates could also show a trend towards larger values. The opposite should occur for an
increased number of intermediate mass stars in the IMF. This is something that can be easily tested
with the population synthesis models.}

\discuss{Cervi\~no}{When you use the TP-AGB tracks you compute a mean value, but there will be
some scatter (TP-AGB stars show variations of a few magnitudes). Have you estimated how large
this scatter is? How do you manage to integrate isochrones when they are a ``band'' and not
a single well defined path? How does the TP-AGB affect the surface brightness fluctuations?}

\discuss{Bruzual}{The evolutionary tracks for TP-AGB stars computed by \cite{PM07} provide well
defined values of $(age,~log~T_{eff},~M_{BOL})$ at all points along the tracks. We use their results
as published. Presumably the aspects you mention have already been taken into consideration
by the authors of this paper in their careful comparison with observations of real TP-AGB stars.
Our isochrones are thus a narrow line and not a ``band''. At the moment I am working with Rosa
Gonz\'alez-L\'opez-Lira in a paper where we examine the influence of the new TP-AGB evolutionary
tracks on the surface brightness fluctuations.}

\discuss{Maraston}{This is not true. There is no scatter.}

\discuss{Bruzual}{Good to hear that.}

\discuss{Maraston}{What differences do you find between the IUE and the HNGSL spectral libraries?}

\discuss{Bruzual}{The signal-to-noise-ratio is considerably higher in the HNGSL data than in the IUE
spectra. As a consequence, the spectral features, absorption lines and discontinuities are much
sharper and better defined. No doubt, the HNGSL is more appropriate than the IUE data to study line
strength indices in the UV}

\discuss{Gustafsson}{You mentioned that your high-resolution indices are quite sensitive to the degree
of degrading due to the stellar velocity dispersion. I suppose that in a real galaxy different types
of stars may have different velocity dispersion. Is this a worry?}

\discuss{Bruzual}{This is certainly a worry when modeling indices like $I_{200}$ and $I_{275}$ defined
by Vazdekis \& Arimoto (1999, ApJ, 525, 144), which are very sensitive to the stellar velocity dispersion.
Models shown in \cite{GB06} assume that all the stars share the same value of the velocity dispersion whereas
in reality on should consider the dynamics of different stellar groups and use the corresponding $\sigma$
for each kind of stars.}

\discuss{Han}{In your second conclusion, you said that there is room for
improvement in UV. We know that UV in old stellar populations is from hot
subdwarfs (EHB). We have a binary model for the formation of hot subdwarfs and
the model is well established for galactic hot subdwarfs. We applied the model
to evolutionary population synthesis study and proposed a binary model for the
UV-upturn of elliptical galaxies (Poster V.10)}

\discuss{Bruzual}{This is certainly an interesting approach. Binary star evolution is missing from most,
if not all, stellar population models. The source of the UV in old populations is not well established yet.
Besides EHB stars in some galaxies there is residual star formation. In my talk I referred
to the improvement that is needed in the UV spectral libraries, but for sure we also need better evolutionary
schemes.}

\end{discussion}

\end{document}